# Holstein polarons and triplet bipolarons with NNN hopping

Monodeep Chakraborty, A. Taraphder, and Mona Berciu

# Holstein Polarons and Triplet Bipolarons with NNN Hopping

Monodeep Chakraborty[1], A. Taraphder[1,2] and Mona Berciu[3,4]

[1]*Department of Physics, Indian Institute of Technology, Kharagpur, India*
[2]*Center for Theoretical Studies, Indian Institute of Technology Kharagpur, India*
[3]*Department of Physics and Astronomy, University of British Columbia, Vancouver, British Columbia, Canada, V6T 1Z1*
[4]*Quantum Matter Institute, University of British Columbia, Vancouver, British Columbia, Canada, V6T 1Z4*

**Abstract.** We study the ground state of 1D Holstein single polaron with next nearest neighbour electron hopping (NNN), employing a variational approximation based on exact diagonalization. Our investigation reveals that, depending upon the sign and magnitude of the NNN hopping integral with respect to nearest neighbour hopping, the polaron band minima may occur at non-zero $k_{GS}$. We compare the present scenario with the SSH polarons, where a similar feature is also observed, albeit, due to very different mechanism. Our initial investigation of triplet bipolarons, in presence of an attractive extended Hubbard interactions, further substantiates the differences between the present model and the SSH model.

## INTRODUCTION

The interplay between fermionic particles and the bosonic degrees of freedom which represents the quantum fluctuations in the external bath, forms a cornerstone to understand many important phenomena in condensed matter physics. In this communication we present our numerical investigation of eletron-phonon interactions within the Holstein paradigm, adding a next nearest neighbour electron hopping term to it. Marchand et. al. [1] reported the occurrence of minima of the single polaron dispersion E(k) at non-zero k ($k_{GS}$) and accompanying sharp transitions beyond a critical λ for the Su-Schrieffer-Heeger (SSH) polarons. The reason for such a transition was the dominance of phonon mediated next nearest neighbor (NNN) hopping in the SSH model. We obtain a similar band minima at non-zero k, for the Holstein polaron with NNN hopping. We compare the results of these two different models and delineate the qualitative reasons for very different behavior of the triplet bipolarons in these two models owing to very different origin of the NNN hopping process.

## THE HAMILTONIAN

The Hamiltonian is given by:

$$H = -t_1 \sum_i \left( c_i^\dagger c_{i+1} + h.c. \right) - t_2 \sum_i \left( c_i^\dagger c_{i+2} + h.c. \right) + g\omega \sum_i n_i \left( b_i^\dagger + b_i \right) + \omega \sum_i b_i^\dagger b_i + U \sum_i n_i n_{i+1}$$

$c_i^+$ and $c_i$ are the electron creation and annihilation operators at the site **i**. $b_i^+$ and $b_i$ are the phonon creation and annihilation operators. ω represents the frequency of the Einstein phonon mode, while $t_1$ and $t_2$ are the hopping strengths of the electron between the nn and nnn sites respectively and **g** denotes the local e–ph coupling. U is the strength of the bare nearest-neighbor (NN) interactions. The variational approximation based on exact diagonalization(VED) [2,3] has proved to be one of the most successful methods to deal with polarons and bipolarons in various electron-phonon (boson) models.

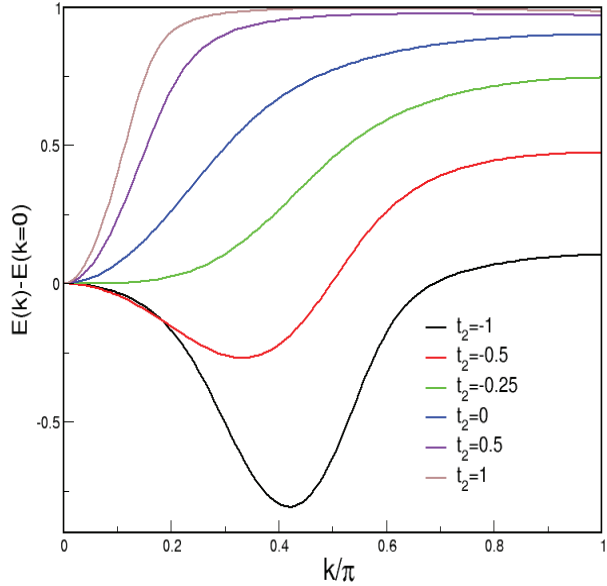

**FIGURE 1.** The polaron band dispersion for different values of $t_2$ for **w** =1 and **g** =1. Energy parameters are in units of $t_1$ = 1.

Here we apply the VED method to obtain the ground state and wavefunction of the single polaron and then go further to calculate a few other quantities like effective mass and average phonon number to comment on the effect of NNN hopping on the Holstein polarons. $t_1$ is set to 1 for all calculations and all other quantities are stated with respect to $t_1$.

## RESULTS

Figure 1 shows the calculated polaron dispersion for oscillator frequency w=1 and e-ph coupling g=1 at different values of $t_2$. We have chosen this particular e-ph regime as this is the intermediate coupling regime where VED is most accurate [2,3]. The polaron band minima remains at k=0 (in units of p) for all positive values of $t_2$ as well up to $t_2 > -0.25$. Here the transition in $k_{GS}$ from zero to non-zero k is entirely due to the sign and magnitude of $t_2$, which explicitly appears in the Hamiltonian and simply mimics the similar transition that occurs in the bare electron dispersion. This behavior seems to be like that noted for SSH polarons, however there the NNN hopping terms do not appear in the bare Hamiltonian and instead are dynamically induced through phonon emission and absorption. As the e-ph coupling becomes dominant above a critical strength depending on the oscillator frequency w, the dynamically generated NNN hopping becomes dominant and drives the transition. It should be noted that in the present model, the value of $t_2$ (=-0.25) below which the $k_{GS}$ assumes non-zero values, is independent of oscillator frequency w and the e-ph coupling strength g; as mentioned above, this is the value where the bare dispersion also exhibits a transition. Hostein model being a g(q) model, the transition of $k_{GS}$ to non-zero values comes from the NNN bare hopping term present in the Hamiltonian as opposed to the g(q,k) SSH model [1] where the NNN hopping arose with the self generated phonon mediated NNN hopping and was a function of w and l. Figure 2(a) shows the calculated $k_{GS}$, (b) the $E_{GS}$, (c) the black solid line gives the ratio of the polaron effective mass with that of a free electron without the NNN hopping ($m_0=1/2t_1$), whereas the red solid line gives the ratio without any e-ph coupling and (d) the average ground state phonon number as functions of NNN hopping strength $t_2$. A sharp transition in $k_{GS}$ occurs at $t_2$ =-0.25 and it goes towards k=0.5 as $t_2$ is further lowered. Otherwise, this transition is very different from that observed for the SSH polarons [1]. A look at the evaluation of effective mass ratio (fig3(c)) with respect to $t_2$, very clearly shows up the entirely different mechanisms at play here as compared to its SSH counterpart. The red solid line in fig2(c) amply substantiates this fact. In SSH model as we increase the e-ph coupling l of the SSH variety at a given oscillator frequency w, the ratio of the effective mass after the transition ($l_c$, where $k_{GS}$ becomes non-zero) quickly flattens out to a steady value as we reach the strong correlation regime. However, for the Holstein polaron, the ratio again increases linearly as the $t_2$ is gradually lowered. Similarly, the average phonon number ($N_{ph}$) decreases after the transition, suggesting a weakening of the phononic activity in the Holstein polarons with gradual lowering of $t_2$. Whereas, in SSH model, a linear increase in $N_{ph}$ is observed, which points at an enhanced phononic activity with increase in l. However, in the present case one needs to keep in mind that change in the bandwidth due to the introduction of NNN hopping changes the e-ph coupling strength which also effects the average phonon numbers. In a nutshell, in the present model, all the features we observe are arising due to the NNN hopping term introduced in the Holstein Hamiltonian, whereas in SSH model, the NNN hopping is a consequence of phonon mediated NNN hopping, entirely a function of l and w. Recently, it has been reported by Sous et. al. [4], that sharp transitions have also been obtained in SSH bipolarons of triplet variety, for whose stable formation an attractive intersite Hubbard U is necessary. They have noted the appearance of a phonon induced repulsion between the two particles, which requires an additional attractive U for a stable bipolaron to form. The exchange of phonons between the particles also mediates a "pair-hopping" term, which moves the pair as a whole. The latter is responsible for the appearance of a sharp transition to a ground-state with a non-zero $k_{GS}$ for the bipolaron depending on the values of U, l and w. Our initial investigation of the present model with two electrons of same spin, reveals that although, here also we need an attractive intersite Hubbard U to stabilize the bipolaron formation, however, we did not find the occurrence of bipolaron band minima at non-zero k.

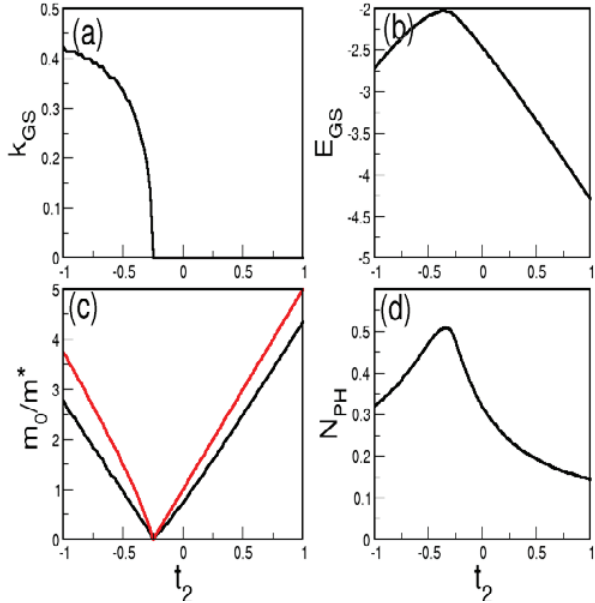

**FIGURE 2.** The calculations are done at **w** =1 and **g** =1. (a) the $k_{GS}$, (b) the ground state energy $E_{GS}$, (c) the black solid line gives the ratio of the free electron effective mass to that of the polaron effective mass, whereas the red solid line gives the same ratio with e-ph coupling and (d) the average round state phonon number ($N_{ph}$).

This clearly shows that phonon-mediated "pair-hopping" terms cannot appear in the g(q) Holstein model. These results strongly support the idea that different e-ph couplings lead not only to different single polaron dispersions but also to (qualitatively) different effective phonon-mediated interactions between the polarons. This should have very significant consequences for the properties of such systems with finite concentrations of particles.

## ACKNOWLEDGMENTS

We appreciates access to the computing facilities of the DST-FIST (phase-II) project installed in the Department of Physics, IIT Kharagpur, India. MB would like to acknowledge the support from NSERC and QMI.